\documentclass[12pt]{article}

\usepackage[paper=a4paper,left=20mm,right=20mm,top=25mm,bottom=25mm]{geometry}
\usepackage{amsmath}
\usepackage{amsthm}
\usepackage{amssymb}
 \usepackage{boondox-cal}

\def\ben{\begin{enumerate}} \def\een{\end{enumerate}}
\def\beq{\begin{equation}} \def\eeq{\end{equation}}
\def\beqn{\begin{equation*}} \def\eeqn{\end{equation*}}
\def\bea{\begin{eqnarray}} \def\eea{\end{eqnarray}}
\def\ba{\begin{array}} \def\ea{\end{array}}
\def\beann{\begin{eqnarray*}} \def\eeann{\end{eqnarray*}}
\def\beasn{\begin{sneqnarray}} \def\eeasn{\end{sneqnarray}}

\def\bi{\begin{itemize}} \def\ei{\end{itemize}} 
\def\bd{\begin{description}} \def\ed{\end{description}}
\def\ids{\item[DS:]}
\def\iat{\item[AT:]}

\def\ea{\'e}

\usepackage{graphicx}
\usepackage{xcolor}
\usepackage{eufrak}

\title{Memories of my early career in relativity physics\footnote{The text presented here and revised by the authors is based on the original oral history interview recorded in Warsaw on June 24 and June 28, 2016. A technical contextual analysis by one of the authors of some of the physics discussed here is in preparation \cite{Salisbury:2019ab}.} }

\author
{Andrzej Trautman$^a$ and Donald Salisbury$^{b,c}$\\
\normalsize{$^a$Institute of Theoretical Physics, University of Warsaw, Warsaw, Poland}\\
\normalsize{$^{b}$Austin College, 900 North Grand Ave, Sherman, Texas 75090, USA}\\
\normalsize{$^{c}$Max Planck Institute for the History of Science,}\\
\normalsize{Boltzmannstrasse 22, 14195 Berlin, Germany}}

\begin{document}

\maketitle

\section{Early childhood and education}

\bd

\ids  Let us start with early history. Can you say a bit about your origins and your early education?

\iat I was six years old and was due to go to primary school when the war broke out in 1939. So I could not go to school and I spent the first two years of the war in the countryside and therefore did not go to any school. I  was just reading. I learned to read quite early, when I was six. So that way I had some, so to speak, education at home and then after three years I returned with my mother to Warsaw. My father had died of heart illness in 1941. He was taken to a hospital in Lublin from Krzczonow, a small  village where we spent, with my mother, the years 1939-42.

 We returned to Warsaw and I started to go to primary school.  I attended essentially  two years of school. Then the Warsaw uprising broke out in 1944.

\ids  Were you and your family directly involved in the Warsaw uprising?

\iat No, none of us. Well my father was dead at that time and my mother was so occupied with me. I was eleven years old. 

\ids In what language did you communicate with the Germans?

\iat Well, we did not communicate at all. At the end of  April 1945, when the front was approaching,
we were forced to walk West, as if the Germans wanted to protect us from the Bolsheviks.
On one occasion we were with a group  of other Poles and  we had nothing to eat. One of them knew some German,
 so they decided they would go to the local authorities and ask for some food. I was taken  along
  to show that there were children with them. We went to the mayor of the city of wherever that was and of course he said, ``no, just go West. Go West, we have no food". These were the final days of the war. Almost everything was in complete disorder. I  didn't  know any German at all, although my father knew some German. He was of German origin as the name indicates. The tradition is that this was an Austrian origin. I think he knew a little German but he did not try to continue it with me. He never made any effort to encourage me to learn German.

My father was a painter, an artist. He taught drawing in high-school. My mother was of French origin. The French connection was stronger. She was born in Spain and she spent some time in France. She knew French rather well. Well, we will come to this.

\ids So what happened then on the trek from Warsaw?

\iat Marshal Rokossovsky  forced us out of Legnica  and we were taken by the Polish authoritiees now
 by car to a town    called Gr\"unberg in German. That was interesting because Legnica, the previous town, was completely empty of Germans. Germans just fled to the West. And this other town, a smaller town, was full of Germans. The Germans did not move out. Apparently it was a question of how the front had moved.  So we were given a room in a house owned and inhabited by the Germans. It was not a very pleasant situation. So my mother decided after a few weeks to go to central Poland, to Lublin, where we had family.  My father's family was there. It was clear that the Warsaw apartment was destroyed. There was no point of returning to Warsaw so we went to Lublin. And, well, things were hard. People gave us a little place to live but it was not a proper place. And my mother had no work, so we really had no income. I started school but after a month my mother decided that we would go to France. I was then twelve years old.

As I said, she was of French origin. My maternal grandfather was French and his wife was Polish. My grandfather was at that time already dead but my maternal grandmother lived in Paris and my mother decided we would go and join her. I do not really know how she did it. She went to the French embassy or consulate in Warsaw and somehow she convinced them. She had of course no documents to support this.  The only thing was that she spoke French, probably good French, and she could give details of her origin.
So she convinced  the French officials that we deserved to be sent to France and that was in September or  October, 1945.

\ids  And you had no identity papers whatsoever?

\iat  Well  my mother had a German Kennkarte. You know the Germans gave Polish people German identity cards, but not stating that we were German, stating that we were Polish.  It is curious how we went to France. You see, the officials at the embassy in Warsaw arranged for us to go to the airport and there was a military French plane on the tarmac  - also with some other Polish people -  and we traveled by that French military plane to France. Much later, my mother had to go to the Polish embassy in Paris and when they heard about the way we reached France they said that that was entirely illegal. They were going to do something about it but probably they did not do anything. In any case we were in Paris and  we got  some help from my grandmother. She was already a quite elderly person and she had also a very small, very modest room -  not really an apartment. It was a hard life. But still my mother finally found some sort of a job -  low paid,  but a job. So  we could live. And I went to school - well - to two schools. The first was a Polish school  for the Polish soldiers  who were scattered in the West. There was a camp of Polish soldiers in La Courtine and they organized a  high school.  After that I went to the Polish school in Paris which was run by the Polish embassy. I got the high school diploma there after two years and then after that we returned to Poland. That was 1949. We decided that it was best. My mother was a real Polish patriot with French roots. And we returned to communist Poland in 1949. Many people, you know, when I tell them the story they really cannot believe it.  It was a bad time.

\ids Did your mother discuss the decision to go back with you?

\iat Well, we discussed it. I also wanted to go back because when I was finishing the high school run by the Polish embassy, I was told that  if I returned to Poland I would be admitted without examination to a university of my choice and given a stipend, and this is what happened.

\ids This played a major role in your mother's decision to return?

\iat This was from my point of view a very good decision.

\ids In your limited high school education and in Paris had you already begun to develop an interest in science?

\iat Yes. Ah, yes, this is a good question. I will comment on this.  I was clearly interested in mathematics and physics at that time in school and towards the end, when I was about to finish the school and make decisions, you know. We had already decided to return to Poland but I had to choose what field to study and where to study. So I asked my teacher, the lady teacher who taught physics. I told her that I was interested in mathematics and physics and that I  hesitated on what I should study. ``Oh", she said. ``Look, I have studied physics and look at what I am.  I am a teacher. And let me tell you this is a very bad profession. I do not encourage you to study either mathematics or physics. You should go to the polytechnical school and become an engineer. That will be a much more interesting, and better paid job." 

 We were returning to Poland from that school - a group of perhaps twelve kids. Since most of them were going to engineering school, I followed them. I was sixteen.

\ids Can you identify what aspect of physics and mathematics attracted you at that time?

\iat Well, I would say theoretical - already at that time - theoretical physics and mathematics as a tool. I read some popular science. I read about relativity and quantum mechanics and that interested me very much. Of course, I did not understand all the words. The ideas, so to speak, appealed to me.

\ids I would expect in that era in 1949 science wasn't that popular and not that much supported - compared to later times which I  had directly experienced where there was more of a societal respect for science, for the study of science.

\iat Yes, science was not respected. So that was my interest and my decisions were not really determined by societal  considerations. The only societal influence was this comment by the teacher!

So I went into perhaps the best technical engineering school in Poland, the Warsaw Technical School, and I went to the electrical department that was again in that school considered to be the best. It was the most physical in a sense. Later it changed and the interesting part became electronics. When I started  the first year was just general electricity but then the second year was already specializing in electronics. So I am  a radio engineer, but you should never ask me to repair a radio.

So well it was clear after the fourth year that this was not what I wanted to study. I was really interested in physics or mathematics and I even tried to study simultaneously engineering and physics, but that was not allowed and it was a wise decision in the sense  that it would have been too hard. Physics requires  labs and engineering also requires working in the labs and they could not be combined. Towards the end of my engineering school I went to study mathematics because that was easy to do and I really finished. I passed almost all the examinations but then I decided there was no point for me to have a second degree.

\ids Is it true that you had wanted early on to switch from engineering to physics but were dissuaded from doing so because you would no longer get state support?

\iat  Yes, there was something like this. This is true but there was another aspect. Perhaps I did not mention that  I was doing this with a friend. We were two of us. We went to see Professor Pie\'nkowski, a  senior professor of experimental physics, in fact the man who built the institute of physics at the University of Warsaw - a very, very distinguished person from this point of view. And he was very kind. We asked if he would be willing to sign some papers that he agrees with
our plan to study simultaneously physics and engineering. He said ``yes, well I respect you, your love of physics and I understand and appreciate it and I will sign whatever you need. But let me tell you that this is  not going to work. This would be  very hard."  Yes, he was a very wise man. So we didn't. We were told by the government that they are not going to allow us to do this and we did not fight any further because we were convinced by the opinion of that senior professor. But later, when I was finishing engineering at Politechnika, there was a lecture course in theoretical physics  given by Jerzy Pleba\'nski, a student of Leopold Infeld.  Well, he was a very, very good lecturer, very outgoing,  extroverted and so on. So I enjoyed his lectures enormously and I would during breaks go and ask him questions and enjoy very much conversations with him. And when he saw that I  was so interested in physics and that perhaps my questions signified that I really have some understanding, perhaps very small, but some knowledge, he said he could supervise me. This was in 1954.

So Pleba\'nski confirmed my interest in theoretical physics and suggested that first of all he would supervise my Master's thesis even though it was in the engineering school. However, my Master's thesis was really in theoretical physics and then he suggested that after I get M.Sc. degree I should go and work in the group of Infeld at the university on general relativity. And this suited me very much, so I accepted.

\ids Wonderful.  But your  thesis, as I think I read, was in radio engineering. Is that correct?

\iat Well, you see, My diploma says  radio engineering because this was the program that I followed and this was the degree that the engineering school was allowed to grant. But my thesis, the piece of work which was in fact later published for which I got the degree was in theoretical physics. It was in fact something proposed not by Pleba\'nski but by Wojciech Rubinowicz. He was another senior Polish physicist. And that was a problem connected with the theorems of uniqueness in the theory of partial - of hyperbolic differential equations. There was a special case that Rubinowicz was interested in and I was asked to consider this case and generalize the well-known theorems for such equations but for this particular case that Rubinowicz  had in mind. And this is what I did. It was a very minor piece of work but still it was my first publication, something that Rubinowicz presented to the academy \cite{Trautman:1955ab}.

So there was really no connection between my actual thesis, the work that I did, and the degree that was granted to me by the engineering school. They were, so to speak, very tolerant. They were keen on establishing relations with theoretical physics and therefore they did not mind.

\ids I see, that did not endanger your state support for your studies?

\iat No, it did not endanger it. Well, of course, if I had really wanted to work as a radio engineer I would probably have had problems because I was not really properly educated.
 This is because during the last two years I was interested in theoretical physics while I should really have specialized and have done some concrete work in radio engineering, which I did not. But I was determined several years before I finished that I would switch to physics.  Well I as I described earlier I made an attempt to study physics but I was discouraged at an early stage. I was discouraged that this would be too hard. So therefore I decided to postpone it and finish the engineering school and then only afterwards move to physics - and this is what I did.

\ids I recall you telling me that you actually managed to take some courses in mathematics.

\iat  Yes during  the third and fourth year at the engineering school I took courses in pure mathematics, but I took and passed the exams of all courses. What remained to get the degree in mathematics was to do a thesis.
But when the time for starting a thesis came I had really decided to go into theoretical physics rather than pure mathematics. Therefore I said that is enough. I mean whatever I learned in pure mathematics - what  I say was pure mathematics - was just differential equations.

\ids Did it involve any reference to Courant and Hilbert?

\iat Well, that is interesting. When I started my studies in the engineering school - that was the time when things were beginning in Poland, you know. They were bringing sometimes books, reprints of books made in Germany and the allied authorities, probably Americans. Among these resources was the reprinted Courant and Hilbert set of volumes.

\ids Right.

\iat I do not know if you have seen this.

\ids No, the reprinting I have not seen.

\iat It was not really a reprint.  It was stated that this was a reprint. And I could buy it in a Polish bookshop in Warsaw for a relatively small amount of money. And in fact I did study it a little - not the entire two volumes. I would not have been able to study it all but I did read - with a dictionary - because my German was poor  - but you know mathematics is never so hard.

\ids  What about books in Russian like Landau and Lifshitz, Gelfand and Naimark or Minlos, etc. The Soviets were very prolific in the 1950's. What literature on Relativity was available at the beginning of your studies? Did you know Bergmann {\it Introduction to the Theory of Relativity} 1942, Synge and Schild {\it Tensor Calculus}, Lichnerowicz {\it Th\'eorie Relativiste de la Gravitation et de l'\'Electromagn\'etisme}  which became very influential in those days and must have been for your work on gravitational radiation. Or his {Global Theory of Connection and Holonomy Groups} which was a revelation for those who wanted to work ``coordinate free"
 - a wonderful introduction into the calculus of differential forms?

\iat In the early days, I had access and studied only the books in Russian, but not only by the Russian authors,
but also the Russian translations from English and French of the books you mention.
\ed

\begin{figure}
\includegraphics[width=\linewidth]{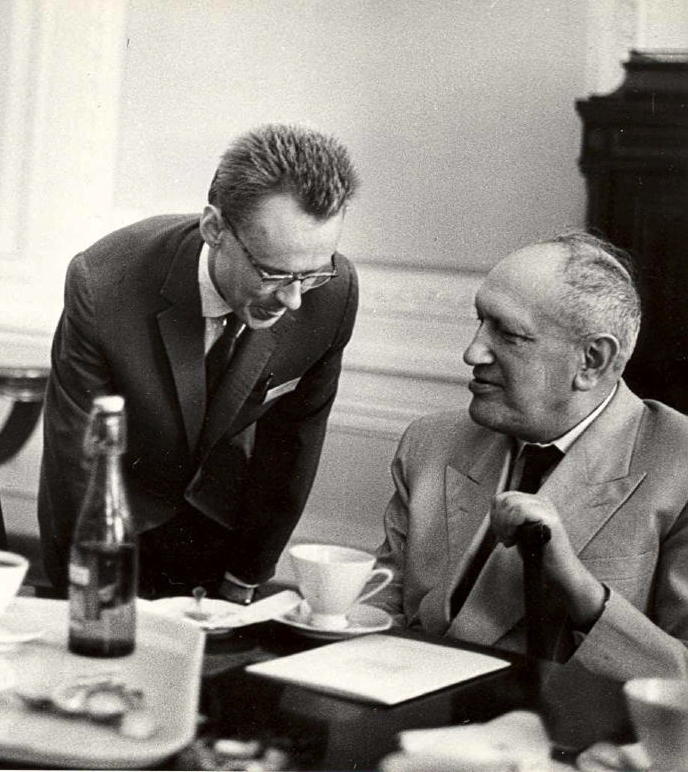}
\caption{Andrzej Trautman and Leopold Infeld, Jablonna, 1962. Photo by Marek Holzman}
\end{figure}

\section{Warsaw 1956 - 1958}

\bd

\ids So Pleba\'nski  proposed that you began a thesis with Infeld...

\iat Yes, that's right. And he proposed what I would work on. He suggested gravitational radiation. But he said: ``Let me warn you. Infeld is against it. He does not believe in gravitational radiation."  With my engineering background I knew quite a lot about electromagnetic waves, and I knew that there was a certain similarity between  the linear approximation to gravitation and  electromagnetism.  So I had the impression that there should be something like gravitational waves. Essentially all of my work with Infeld consisted in finding various arguments, doing  calculations and estimates in order to show that there can be gravitational radiation. I really did all my early work on waves alone, without  any input from Infeld.

Infeld did not accept my arguments. He was already in a rather poor shape and the only thing that he could use and understand was the EIH method which is not really suitable for radiation because it is a slow-motion approximation. Therefore I had a hard time. But on the other hand Infeld showed character.  He saw that I was doing some good mathematical work when searching for waves and therefore at the end he agreed to accept what  I had: a few publications  and some manuscripts. He accepted it all as my PhD thesis even though he did not agree with the conclusions. He had to write an assessment and he did.  He was very fair and I cannot really complain.  He did help me even in other ways. One very important thing was he arranged for me to get an apartment in Warsaw. I lived outside of Warsaw in something that was really low quality. There was not even running water. It was really very bad and he arranged for me to
be given  one of those apartments built under communism.  It had running water, heating, a toilet, a bathroom. It was in a sense entirely modern but not of very high quality. The walls were thin and therefore there was a lot of noise coming from the neighbors, but when you compare to what I had before,  it was really luxury. What Infeld did for me  was not at all easy at all in those days.

\ids Before we get into the content of your thesis perhaps you could say a bit more about Infeld himself. What kind of privileges did he enjoy?

\iat  Yes, he had a strong position. For example,  he had a telephone that was part of a special exchange
allowing him to reach important people, even the Prime Minister.  He did not really use this very much, but occasionally when he wanted something important to arrange, he would. This is one example. Another one is that he had a car with a {\it chauffeur} which was part of the services of the Prime Minister.
One of the cars that were at the disposal of the Prime Minister was granted, with a {\it chauffeur}  to Infeld.  He used it very much, going  from home to work. He would be driven, which was in fact bad because he did not get much exercise.
Bad for him but still - convenient.

\ids Infeld of course had many contacts from his earlier research positions abroad. He did play an important role in encouraging international intercommunication among relativists.

\iat Exactly. He was always repeating that it is very important that we have contacts, that we invite people from all over the world to Warsaw and that we send people to postdoctoral studies.
He insisted that  young people should be sent to good centers of research,  but only after the doctorate. In this respect,
 he made an exception for me.

\ids Do you remember who were the foreigners invited to visit Warsaw?  Were there any contacts with relativists in the West  or with Vladimir Fock's people in the Soviet Union?

\iat Surprisingly, at the time of my early work, there were not even Russian visitors to Warsaw working on general relativity theory.
One day Infeld went to Moscow and Dubna with a group of Polish physicists, but I was not taken along.
The first Western visitors working in general relativity were Felix Pirani and Ivor Robinson. Later, starting in the 1960s,
contacts with Western scientists became easier.

\ids Would you say some more about Felix Pirani?

\iat Infeld and Pirani knew other from Canada, from Toronto. Felix was a younger person, but they were on good terms. Felix  was at that time in London and Infeld invited him to spend  several weeks in Warsaw.  And he came here in 1957 and it was really very exciting for me because a  few months before his visit I read his paper. 
The title was `Invariant formulation of gravitational radiation theory' \cite{Pirani:1957aa}. He said that one should look at the Riemann tensor   in order to characterize what to consider as gravitation radiation. And I read this paper with great interest and in fact had some comments on it which I presented to Felix when he came.

\ids So it was you who suggested the asymptotic behavior?

\iat Well I think I was the first to suggest that radiation is present asymptotically when the one over $r$ term is of type N  in the Petrov classification. But  I think I was merely  saying that it has the same algebraic structure as a plane wave - but only the one over $r$ part, not the rest.
He of course was so impressed because he understood much. He felt that saying that the Riemann tensor has to be everywhere of  the special type is not a good idea, but he didn't know how to improve it. And so he was  impressed
by my suggestions.

\ids But he had not seen your papers before he arrived in Warsaw?

\iat No he had not seen my papers. 
They were published in the {\it Bulletin of the Polish Academy of Sciences}. Infeld was a patriot. He was member of the Academy of Sciences and he in fact contributed to the establishment of the {\it Bulletin}. He published his papers there and he made me publish most of my early papers  there.
He said ``You should publish there because this journal has to have a good standing." As a result of this, those papers  reached hardly anybody.  But Felix was so impressed that when he returned he invited me to go to London and the following year, in 1958, I went there.

\ids You had already encountered something analogous to the Sommerfeld condition in your studies in radio engineering. Or were you involved in radar?

\iat  I must have gotten the Sommerfeld radiation conditions from Courant Hilbert, not from the engineers. 
The engineers would not have known.  But towards the end of my studies in the engineering school, I gave a lecture course. You know, I was considered to be sufficiently good to lecture even though I had no degree. So I lectured on electromagnetic waves.

 I would like to say that I have  positive  reminiscences from the engineering school. They were really very tolerant, open-minded people. They were not closed in their specialty. That was a good school. Well, there was a senior distiguished person, Professor Janusz Groszkowski.  He was a specialist in vacuum tubes. He was a world authority in vacuum tubes and wrote books on the subject. He later became president of the Polish Academy of Sciences and I had very good contact with him. I must say, even though the  engineers would not know Lorentz transformations and things like that, but there was a lot of tolerance for physics.
 
\ids Perhaps you could say something more about equations of motion using the EIH method and the relation to gravitational radiation. 

\iat I had to work with EIH. Infeld knew that  introducing the radiation terms in the EIH method required going to high order.  Then his claim was that these terms can be eliminated by coordinate transformations. In the EIH method the coordinate transformations act a little like gauge transformations in electromagnetism. And in fact he had some sort of a proof that this is so, but he did it, so to speak, component by component. When you use the EIH method,  there is a natural split of the metric tensor into three parts, the time-time components, time-space components and the space-space components. And they  transform separately under those gauge transformations and Infeld  showed that one can first eliminate the radiative term in the time-time component, then in the space-time component and so on, but separately. Then it turned out -  it was showed for the first time by Josh Goldberg \cite{Goldberg:1955aa} - that one cannot do it at the same time for all the components. The criterion that  Josh gave was that one has to take these three parts and make out of them the appropriate component of the curvature tensor, but you need for this all three of them. You cannot do it separately. And then you look at the curvature tensor. If it is non-zero, then you can not eliminate at the same time all three radiative parts.  Infeld was in such a poor shape that even though  after reading the paper by Josh  I tried to   explain this to him,  he could not accept the argument.
\ids That paper appeared in 1955.
And you had access to the {\it Physical Review} then?
\iat Yes.
\ids So  why do you think Infeld had this gut feeling that radiation should not exist?
\iat My guess is that this is connected with his being in Princeton in 1936 and '37 when Einstein and Rosen were there and they wrote their  paper that claimed that gravitational waves did not exist \cite{Einstein:1937aa}.
The Einstein-Rosen paper  was rejected by Robertson who refereed it for {\it Physical Review} as we now know.  Infeld had of course interactions with Einstein and he knew about that paper.  Einstein was  reluctant to accept  Robertson's criticism or explanation.  I have the impression that Einstein for quite some time still had doubts about waves. Following Robertson's
suggestion, he  wrote  a paper presenting the waves as cylindrical,  but  he still had the impression that plane waves do not exist and he might  have said  that if there are no plane waves, then there  cannot be proper radiation. Something like this. So I have the impression that Infeld was under Einstein's influence.
  Robertson rightly said that the difficulty that Einstein and Rosen encountered was a question of coordinates, but Einstein was convinced  that he was using the right coordinate system,  adapted to the symmetry of the problem
   under consideration. So it is not quite clear to me what was really Einstein's attitude later. In any case Infeld's attitude did not change until toward the end of his life. He did not believe in gravitational waves until he was finally convinced of the existence of gravitational radiation by the work done by him with my
wife, R\'o\.za Michalska-Trautman \cite{Infeld:1969aa,Infeld:1969ab}. She showed him how to correctly account for such radiation within the EIH method. 
   
\ids It is interesting what you point out:  Einstein's belief that plane gravitational waves did not exist  probably played a role  also in Infeld's view of the question. I recall reading that you were impressed not only by Josh Goldberg's paper but also by Ivor Robinson's  demonstration that actually plane waves  did exist.

Continuing with Infeld's attitude toward gravitational radiation, I wonder if you could comment on his collaboration with Plebanski that culminated with their book {\it Motion and Relativity }\cite{Infeld:1960aa}. I notice that they had  acknowledged you in their introduction for ``many critical remarks".

\iat As far as I know the story of the Infeld - Pleba\'nski collaboration on their book is essentially as follows.
At some point, they agreed to write together a book on equations of motion in GRT.  They never discussed in detail how to
organize the book and how to divide this task among them. That project was started after they completed
working on `good' delta functions (not good at all in my opinion: they are not defined, as the ordinary delta functions are,
 as linear functionals on a suitably space of test functions). Those functions were supposed to provide a good description of
 point-like bodies, sources of the field, and help to resolve the problems with singularities inherent with work
with such sources. Infeld wanted to use them in the book.  Pleba\'nski became disenchanted with delta functions in GRT
and, shortly before leaving for Mexico for good, he wrote his version of the book, using a continuous distribution of matter, similar to that used by Vladimir Fock.   Infeld  strongly disliked both Fock and his approach to the problem of motion.  This may have contributed to his rejection of
Jerzy Pleba\'nski's text. Infeld wrote the book by himself, not telling Pleba\'nski about that decision and completely disregarding what Jerzy wrote.

At some stage, Infeld showed me parts of the book that he was writing and asked to check the calculations, but we
never discussed any questions of principle. In fact, I was not interested in the book as I did not like the EIH method
which is based on the assumption of slow motion of bodies and, therefore, not well suited for questions of gravitational radiation.
My only contribution to the book was to prepare a suitable  bibliography.

\ids Do you think this episode contributed to Pleba\'nski's decision to leave Poland?

\iat In my opinion Pleba\'nski left for Mexico because he felt he would not be able to obtain in Warsaw a suitable
research position. Also in those days salaries in Poland were very low and many able people accepted better paid jobs abroad.

\ids Continuing with Infeld, can you say anything about his relation to Myron Mathisson and J\'ozef Luba\'nski?

\iat Infeld never mentioned Mathisson or Luba\'nski to me. I first  learned about Mathisson from my colleague Wlodzimierz Tulczyjew
and I was so impressed by his work that, much later, I organized in 2007 in Warsaw a small conference devoted to him.

\section{King's College, London, 1958}

\ids So Pirani arranged for you to be invited to give a series of lectures at King's College, London, in May-June, 1958.

\iat Yes, by then I already had the essential results. I do not recall whether the papers were out. They are dated '58,  but I do not recall whether when I was going to London they were already printed. But in any case I had the manuscripts and I lectured about the use of radiation conditions in gravitation,  and about those two significant results from those papers:  the loss of energy and the asymptotic behavior.  I am really amazed how I managed to lecture. My knowledge of English was extremely poor.  I think I had only one year of English in high-school.   Those years when I was in high-school, they were the worst years of Stalinism. English was not really the thing that Russians much supported.
 Before I left to lecture at King's I took some private lessons of English. I had a tutor who tried to teach me. When at the
 end of this tutoring I asked him  whether I would manage  to give those lectures, he answered that I would have to try very hard. This really shows how tolerant Bondi and Pirani and other people were.  They  came to all my talks.
\ids You could have lectured in French, couldn't you then?
\iat I could have lectured in French, yes, I know French quite well. In fact with Felix I conversed sometimes in French. He knew a little French, but not  very much. His French was probably only slightly better than my English.
 But still somehow I managed in the sense that people came. They did not abandon me. Later I wrote down the lectures and Felix helped me very much  with the written English.
\ids Those notes became very famous at the time. 
Fortunately for all of us they were finally published \cite{Trautman:2002aa}.

\iat  During that visit, I went for a day to Cambridge,  essentially to see Dennis Sciama;  Roger Penrose came to see Dennis just for  half an hour. So I met Roger Penrose during that visit. But Roger does not remember that. In London I met Ivor who   came to one of my  lectures from  Wales because he had then  an appointment in Aberystwyth. I would say it is interesting  there was something like love at first sight between Robinson and myself.

\ed

\section{Collaboration with Ivor Robinson}

\bd

\iat There was a strong  scientific affinity between Ivor Robinson and myself. I  cannot really explain why, but we understood each other so quickly. From that point on we kept in contact.
Much later, when I visited  Ivor in Dallas, he told me that he considered me to be his younger brother.
I liked and enjoyed this much, especially since I never had any siblings of my own.

So we understood each other very well. But I consider him to be  my teacher, my superior, and I was extremely impressed by the  idea of shear-free null geodesics. I consider it to be his greatest discovery and contribution which unfortunately very few people recognize. It was highly appreciated later, for example in Jordan, Ehlers and Sachs \cite{Jordan:1961aa,Jordan:2009aa,Jordan:1961ab,Jordan:2013aa} and by Kundt \cite{Jordan:1961ac,Jordan:2014aa}

\ids   Yeah, actually this is one thing I wanted to ask you about because there is a reference  in one of your papers  of his having presented  these ideas at Royaumont in 1959?  I have seen at least in two different locations reference to a paper which Ivor gave at Royaumont, but the paper does not appear in the proceedings.
Can you say a bit more on Royaumont? Have you been there?
\iat  Well,  I think he presented these ideas even before Royaumont because he would go to Paris, Hamburg - where there was a major group you know - and of course to England – Cambridge, Oxford. And he would speak and one gets the impression that in some cases people would use his ideas without citing.
\ids Would you happen to recall the title of Robinson's presentation at Royaumont - or at least the subject matter? 

\iat It is my impression that Ivor was the discoverer of the notion of shear-free null geodesic congruences and
of their relation to null (optical) solutions of Maxwell's equations. It is hard to prove this to be true because he was publishing little and late. He would not write, you know.

Being so impressed, when I returned to Warsaw I asked Infeld  to invite Ivor to Warsaw and he came the following year. That is to say  in 1959, in the spring.  And of course he lectured. We started collaborating and he presented a seminar on a solution of the  Einstein-Maxwell field.  Infeld asked him to submit the paper to the Bulletin and it was published \cite{Robinson:1959aa}. 
This is Ivor’s first publication in which he is the only author. The same year there appeared the paper by Ivor, Bondi and Pirani on gravitational plane waves \cite{Bondi:1959aa}.

\ids   Can you talk a bit more about your collaboration with Ivor? 
Well, actually I am of course interested in the physical content of the work which the two of you did together. But maybe you could give just a little more background about his personality, his manner of interacting.

\iat  Well he very strongly needed people to talk to and this was the way he could really work. The sad thing is that it was tough getting a job at a place where there would be people, so to speak, equal to him and that was one of the reasons why he travelled so much. He would present his work because in this way he had contacts with good people. And he did not like to write. Well, he wrote himself our first joint paper,  on spherical waves \cite{Robinson:1960aa}. 

Perhaps let me say a few more words. I think when we started he was in Warsaw but then that was in the spring '59 but not much was done. But then in the fall of '59 I went again to England to Imperial College to spend a year, more than a year with Abdus Salam.
Ivor at the same time went to Chapel Hill, and we continued our collaboration by correspondence. I even have some of his letters.
  And then this first paper was written by Ivor and sent to {\it Physical Review Letters}.
This article was later the starting point of the paper by Jordan, Ehlers, Sachs that I cited earlier.
\ids   Can you say  what the origin of this paper was, what got the two of you thinking about this approach to exact solutions.
\iat  The plane waves were very special. We wanted to have some outgoing waves, radiation from an isolated system.
\ids   So the impetus, initially, was to find an exact solution that manifested gravitational radiation?
\iat  We wanted to find very simple outgoing gravitational radiation. And  Ivor said let us try to use a shear-free congruence of null geodesics which is non-twisting (hypersurface-orthogonal) but diverging. The plane waves are
non-twisting, but have no divergence. The essential change was that one introduces divergence. And then  the equations are more complicated than in the plane wave case but still they could be solved. We didn’t really find the most general solution but we found many concrete solutions.   And then we continued this work in 1961 when I went to Syracuse. Ivor was the first to get a position at Syracuse and then I was also invited by Peter Bergmann. I was due to come in the fall of 1960, but  there were problems with the visa so that I arrived sometime in March of 1961. We then  wrote  our second paper which was essentially devoted to the same topic.

\ed

\begin{figure}
\includegraphics[width=\linewidth]{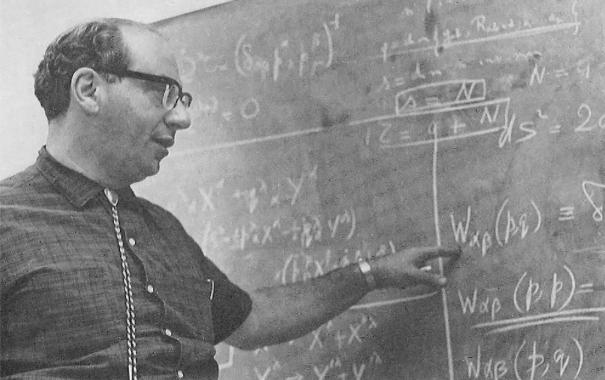}
\caption{Ivor Robinson. Photo courtesy of Joanna Robinson.}
\end{figure}

\section{Syracuse in 1961}

\bd
\iat This was my first visit to the United States and I met lots of interesting people.  R\'oza, my fianc\'e
and future wife was very kindly invited  by Peter to Syracuse   so that we spent  three months together in Syracuse. It was great and  we married shortly after we returned.
\ids   So that was a period of intense activity - extensive interactions of relativists -  that gathering of you folks in Syracuse.
\iat  Yes, that was a very, very good year when there were  Ivor and  Engelbert Sch\"ucking, who became  a very good, close friend.
Art Komar was there. There were a few other people. There were people coming for visits. I think that Ray Sachs came, but I had met Ray Sachs quite early in my  life. I do not remember whether this was in London, England or in Syracuse. I think it was rather in Syracuse.
\ids   Can you tell me a little bit about how this visit to Syracuse came about. You received an invitation from Peter Bergmann?
\iat  Well, I met Peter in London. The only thing is I am trying to remember  whether this was during my first visit - the 1958 visit - or during this second when I was at Imperial College.  I would go almost every day to be seminars, lectures and so on, but I would also spend quite a lot of time at King’s College. My research so to speak was at King’s and I think that Peter Bergmann visited King’s during that second visit.
  And again we interacted and probably Felix Pirani must have   told him something nice about me so that very soon Peter said  ``You must come to Syracuse," and then he arranged an invitation for me. The invitation came quite soon but the problem was with the visa. You see, I applied for a visa but I was a member of the Polish United Workers Party which was a communist party and therefore I had to wait quite a long time.
\ids   But eventually you succeeded in getting a visa. That is a remarkable achievement. How did that happen?
\iat  Well you see, apparently a number of people like Ivor and Bondi and Bergmann must have given to the immigration authorities a letter of recommendation that I was not that bad. Perhaps it was John Wheeler.
 I am sure I must have met John Wheeler at least at Royaumont if not before. John Wheeler may have also said something in my favor – he was perhaps the most influential person.
\ids   In terms of political influence?
\iat  Yes, he was the most important. But it took time for me to get the visa.
\ids   Did Peter Bergmann express interest in any particular aspect of your current research that he wanted to continue discussion with you?
\iat  I would say that he was interested in gravitational radiation. But he was very kind and wanted to  help people working behind the Iron Curtain. Engelbert Sch\"ucking characterized him as being a combination of Einstein and a Saint.
I don't recall which Saint.
\ids   I talked to Engelbert about Peter several years ago. I do not believe he cited the same characteristics but of course he was effusive in his support of Peter.
\iat  I remember when he said this. Josh organized a conference in the memory of Peter Bergmann in 2003.
\ids  I was there. I remember your talk.
\iat  I see, well I think that we discussed this with Engelbert and so many people came, really very many people, and this is when Engelbert said: ``You see, Peter was a combination of Einstein and a saint..."
\ids   Oh, in his after-dinner talk!
 I have this image of Peter that I think is shared by all of his students ... I always thought of him as a gentleman of science. His manner of interacting with people was non-confrontational. He was a warm personality and he was of course genuinely interested and involved.
\iat  Yes, let me illustrate this with an example if you wouldn’t mind: My future wife, then fiancé, came to Syracuse with me. She was a very shy person, you know, and she was asked to give a talk because she was a competent scientist. But of course she was very shy. It was the first time that she would give a talk in English in front of this distinguished group – and there were people like Art Komar who would, you know, always interrupt, ask questions,  or explain this or that. Peter said, lowering voice, ``shut up".
\ids   He did? Oh that’s very unusual. That’s an incongruity in personality that I always  found  striking. But they were very close, and yet they were really distinct personalities. I had never heard that story.
\iat  That shows you that he knew what the problem was, what dangers were involved.
\ids   That’s a good story. There was much going on in Syracuse at this time.
\iat  Oh yes. I mentioned Roger Penrose was there.
\ids   Right and do you recall how  day-to-day interactions occurred amongst all you visitors?
\iat  Well there was also a group called the Kibbutz, well part of it was Jewish of course, Ivor and Art Komar, Mel Schwarz – was there anyone else?
 The idea was that we would meet and have dinner every night, but in different places. But it turned out to be mostly in Ivor’s apartment.
 We were in the same house but there were apartments on two floors. Ivor was on the ground floor and I was on the first floor.
\ids   And so you would get together every evening.
\iat  Someone had to cook. And of course often we would go to a restaurant.
\ids  I guess the discussions were far ranging?
\iat  Far ranging: Politics and science and politics and Jewish problems and so on. It was very pleasant.
\ids   And Peter, while he was in Syracuse, was a part of this?
\iat  Well I think he occasionally came but he was very busy and he also worked extremely hard because he would come from New York City by plane and he had a tiny room of some sort. He could spend two or three nights  with lecturing, so he was tired. So I don’t think he would come very often at that time.

\ids   Do you remember what were roughly the central physics themes that people were involved with at the time? 
\iat  Well, Roger Penrose was of course with spinors, not yet twistors or something. They were writing the first drafts of the book that became later Penrose and Rindler \cite{Penrose:1986aa,Penrose:1988aa}. 

\ids Good, let us return to this later. But first, perhaps you could complete the discussion of your collaboration in Syracuse with Ivor Robinson.

\iat  The title of the second paper was almost identical with the first, but it was a  much longer paper \cite{Robinson:1962aa}.  In that  paper my contribution is more pronounced and I described there the analogous solutions in electromagnetism. I could also explain some of the singularities which appear in  these spherical gravitational waves.

\ids   Yes, I recall that you cite Peter Bergmann  as suggesting that because there exists a singular line in the solutions, this line could be interpreted as being due to incoming matter that would compensate for the outgoing gravitational energy.
\iat  Yes, that’s right.
Well you see the electromagnetic analogue is the following: Suppose that you take a paraboloid of revolution made of a perfect conductor and suppose that you arrange for a plane wave to fall on this paraboloid  in the direction of its axis. This will be reflected  and the  reflected wave will be as if it was coming from the focus of the paraboloid.
 It’s a very simple picture. And essentially this reflected wave is a spherical wave.
\ids   Is that the picture that you had at that time?
\iat  Yes, that I describe in the 1961 paper. But  in the gravitational case we have no paraboloid. We have only singularity. But then suppose that you shrink the paraboloid in the electromagnetic case  so as to become a line and this is where the singularity occurs. Essentially the singularity in the electromagnetic case comes from shrinking the paraboloid to a line. I would not say the same is true for gravitation because we do not have a perfect conductor or perfect reflector.  But as far as the geometry is concerned the analogy is very good.
\ids   Did that come from Peter then?
\iat  No.  That was mine. Peter’s idea was that there may be some matter flowing from infinity. Which is possible, but we did not pursue this idea.

\ed

\section{Jablonna - 1962}

\bd
\ids Could you comment on the significance of the Jablonna meeting in 1962.

\iat An International Conference on General Relativity and Gravitation was held in July of 1962 in Jablonna, a small town near Warsaw. It was organized by Leopold Infeld and his students and collaborators, including myself, Stan Bazanski,
R\'o\.za Michalska-Trautman, Barbara Pierzchalska-Tulczyjew, Jerzy Pleba\'nski, Joanna Ryten,  and Wlodzimierz Tulczyjew. Over one hundred physicists attended the conference. 
There were among them many distinguished scientists such as Peter G.  Bergmann, Hermann Bondi, Bryce DeWitt, Paul Dirac, Richard Feynman, Vladimir Fock, Vitali Ginzburg, Joshua Goldberg, Dmitri Ivanenko, André Lichnerowicz, Stanley Mandelstam, Charles Misner, Christian M\o ller, Roger Penrose, Ivor Robinson, Nathan Rosen, L\'eon Rosenfeld, Ray Sachs, Leonard Schiff, Alfred Schild, Dennis Sciama, Leonard Schiff, Engelbert Sch\"ucking, John Stachel, John Lighton Synge,  Joseph Weber and John Archibald Wheeler.
It was the first large gathering of physicists in Poland after the 2nd World War. Among the participants there were
34 from the USA and Canada, 13 British, 10 from the Soviet Union, 10 from the East and West  Germanies,  a few from
Austria, Australia, Belgium, Bulgaria, Denmark, France, Hungary, Ireland, Iceland, Israel, Italy,  Romania, Switzerland, Sweden, Tunesia and a moderately large crowd of Poles. The conference was significant because it provided an opportunity that was rare at that time
for a meeting of physicists from the free West and the Soviet dominated East. Particularly worthy of attention was the participation of Vladimir Fock and his group from Moscow.
The proceedings were intense. There were 18 plenary expositions, given by Synge, Mandelstam, M\o ller, Pleba\'nski, Ginzburg, Schiff, Sachs, Robinson and Trautman,  Bondi,  DeWitt, Bergmann, Dirac, Lichnerowicz, Misner, Feynman, Wheeler and Fock. There were
25 short contributions, presented at seminars.
The texts of the plenary lectures, with the exception of that by Feynman were published in the volume of the proceedings of the conference \cite{Infeld:1964aa}. The text of Feynman's lecture became available too late to be included in the proceedings.  It was published
as an article in Acta Physica Polonica \cite{Feynman:1963aa}.

\ids Peter Havas claimed that Infeld refused to include equations of motion in Jablonna meeting because
Fock would be there, insisting that this problem had been solved. It appears that he did
this to avoid any discussion of alternatives to the EIH approach, including that of Fock,
who was present at the conference. Is this your recollection?

\iat I remember that Infeld refused to let Havas speak on equations of motion at the 1962
Conference. I think he was reluctant to have as speakers people who did not share his
approach to the problem of motion in GRT.

\ed

\begin{figure}
\includegraphics[width=\linewidth]{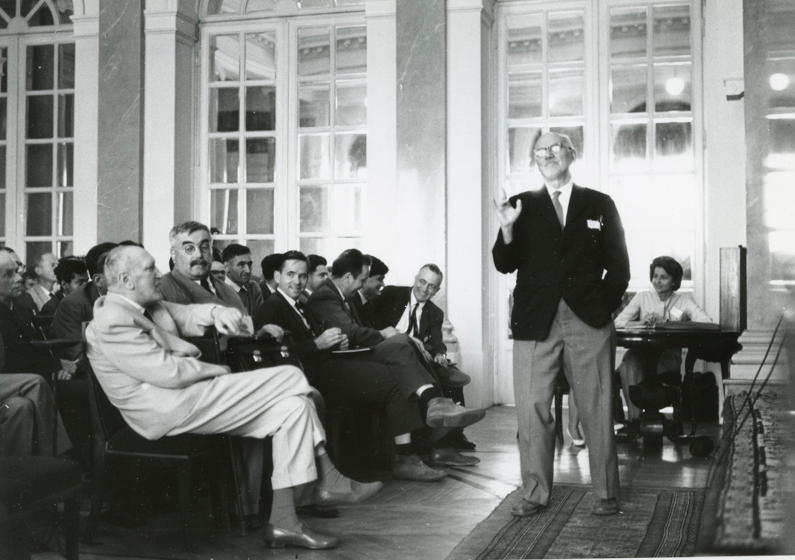}
\caption{In the front row from the left: Leopold Infeld, Vladimir Fock. Standing is John Lighton Synge and seated on the right R\'o\.za Michalska-Trautman, Jablonna, 1962. Photo by Marek Holzman}
\end{figure}

\begin{figure}
\includegraphics[width=\linewidth]{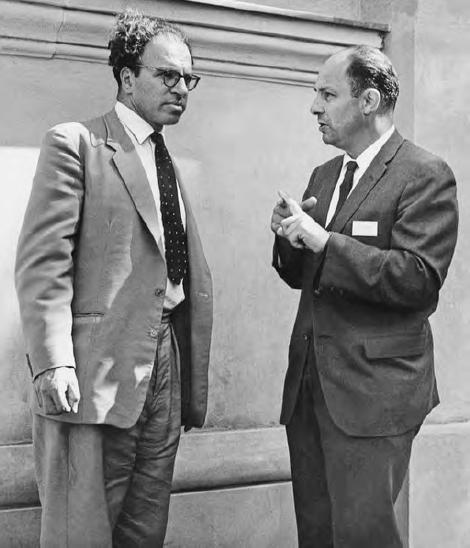}
\caption{Hermann Bondi and Peter Bergmann, Jablonna, 1962. Photo by Marek Holzman}
\end{figure}

\begin{figure}
\includegraphics[width=\linewidth]{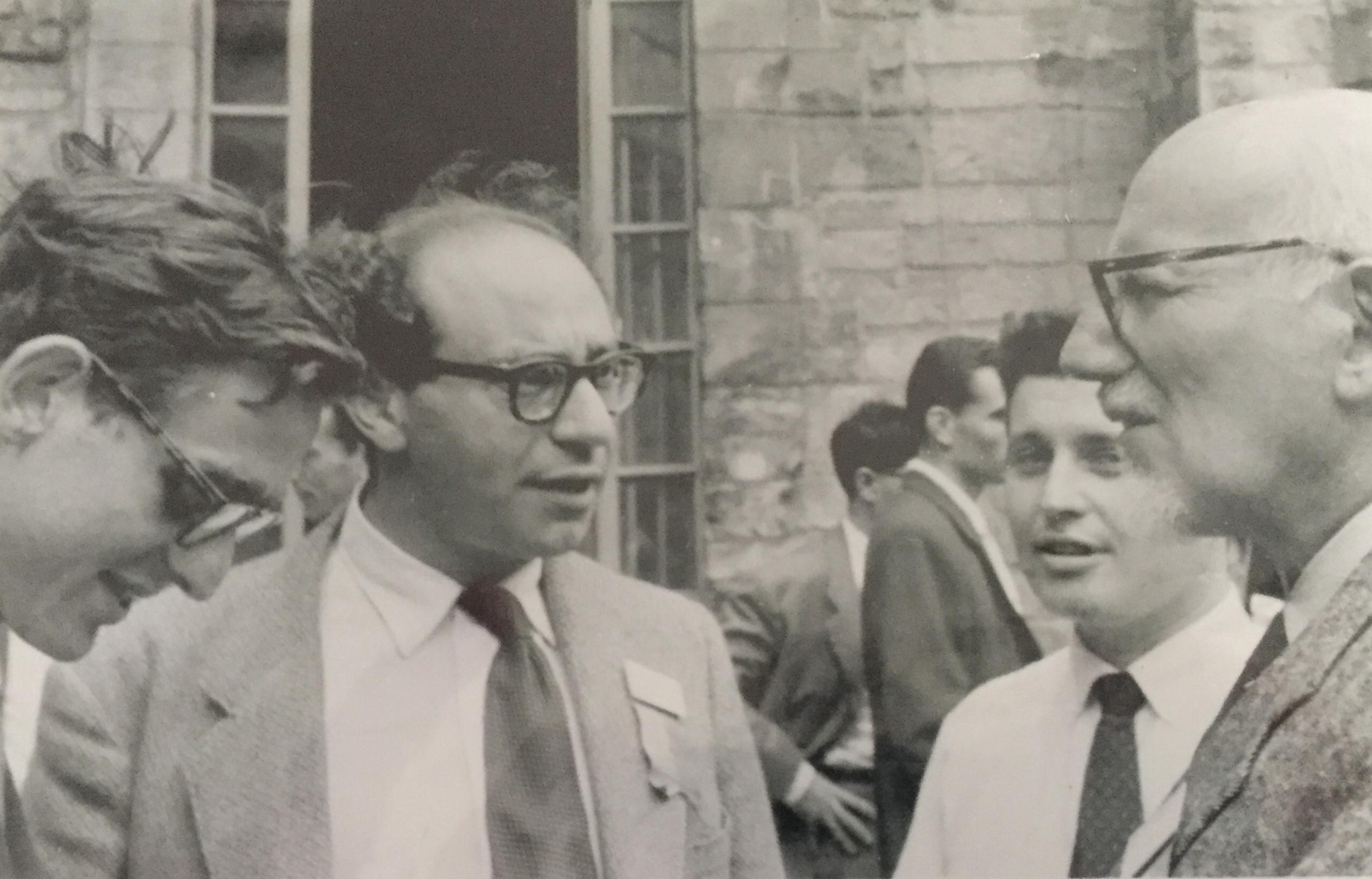}
\caption{Ray Sachs, Ivor Robinson, Art Komar and John Lighton Synge, Jablonna, 1962. Photo courtesy of Joanna Robinson, by Marek Holzman}
\end{figure}

\section{Mathematics, spinors, and fiber bundles}

\bd
\ids  You mentioned Roger Penrose's work on spinors when he was in Syracuse in 1961.
Continuing perhaps a little bit of the discussion about the spinorial approaches, your first exposure to spinorial ideas I guess was probably on your second visit to London? Penrose lectured 
on spinorial approaches to gravity and space-time classification?
\iat  Yes, I was of course exposed to Roger’s lectures and I absorbed some of it but I did not really pursue it.
\ids   You did not immediately see the usefulness of spinorial approaches to classification?
\iat  Yes, but you see I had the impression that this was already done by other people.
 What interested me, but again this interest was not shared by other people, was the global topological aspect of spin structures. In fact  I differed with Roger because I recognized from very early on the validity of the mathematicians insistence on  considering defining spinors on manifolds in terms of spin structure, and that there is sometimes an obstruction. One cannot do it whereas a lot of them did not appreciate this. 
\ids   Did you have that conversation with him about this observation?
\iat  Well, I did, but you see he did not accept spin structures. But you see, Roger is a genius and it is interesting here. He is a mathematician by education. He has a degree in mathematics. But, interestingly enough, he had not had a very good education in modern mathematics.
\ids   I am surprised to hear that. I thought that was his formation – in modern mathematics. When you say modern mathematics what are you thinking of?
\iat  By modern mathematics I mean topology.
 You know, for example, fiber bundles.   Very early on  I had used the  fiber bundle language, but he would not. If you look up his early papers there are no fiber bundles.

\ids You did a lot to promote fiber bundles. Didn't you also clarify as one of the first the modern concept of gauge theories?

\iat  But let me repeat. I consider Roger to be a genius, by orders of magnitude superior to me. Therefore, when I make these critical remarks they are not intended ...
\ids   I understand. No one is going to interpret that as anything other than a side remark. 

Well actually this theme is related to a question I wanted to ask about your interaction with people in Syracuse, with Peter Bergmann in particular. I think it is after you were in Syracuse for the first time in 1961 - I guess in 1967 - when you began to develop an interest in fiber bundle approaches to gauge theories.
\iat  In 1970 I wrote a paper entitled {\it Fiber bundles associated with space-time} \cite{Trautman:1970aa}. 

\ids This became a frequently cited lecture at the time. Some think you were the first to have explained gauge fields using fiber bundles. What can you say about that? 

\iat  Well, this was not the beginning. It was already a mature paper in a sense. 

\ids  Yes, you had already begun in 1967 to look at invariance properties of Lagrangians in terms of fiber bundles.
\iat  Yes, there is a 1967 paper {\it Noether equations and conservation laws} \cite{Trautman:1967ab}.
And the 1970 paper was based in part on a lecture course I gave in Syracuse in 1967.\footnote{A footnote explains that the published paper was an ``expanded and modified version of the notes by H. P. K\"unzle and P. Szekeres of the lectures given at King's College, London, in September, 1967}
\ids   Good, this ties in very well with a question I wanted to ask you about your interaction with the Syracuse group. My own recollection of my graduate days in Syracuse which actually began not long after that. I first arrived in Syracuse in 1968 and then I was gone for two years before I returned. My own background, which I learned through Peter Bergmann and through others at Syracuse, was very much in coordinate-based approaches to general relativity.
And I know that   you have had strong objections to that approach.  I am wondering what kinds of conversations you had with the Syracuse folks in trying to promote your view.
\iat  I tried but not really very successfully. Let me tell you that after I finished this course of lectures at Syracuse which later became the paper – I do not remember but it was probably a couple of months later - Josh Goldberg came to me and said: ``Your lectures were very nice but I am completely sure that fiber bundles will never be of any use in physics."   I saw more of Josh than Peter Bergmann and he came to all my lectures.
\ids   Oh my goodness!
\iat  Well you see it did not take a long time. I think in 1971 or thereabouts there was a conference in Berlin where we again met and Josh came and said  ``I apologize. You were right about fiber bundles.'' That was the time when magnetic monopoles and various instantons had evolved.
So I had these ideas quite early. Well, since the lectures were in 1967 I had them probably in the early '60s to mid-'60s. You know, when you give a course of lectures then you have already been thinking about it for some time.
\ids   Can you recall  what was the inspiration for your belief in the applicability and importance of fiber bundles?
\iat  Well, you see, I saw that just like groups they are natural for  organizing symmetries. Fiber bundles are essentially a natural idea for organizing fields. 
\ids   When did you first encounter them?
\iat  Well, in fact I should also mention  my colleague Wlodzimierz Tulczyjew.
 He was a person who was very enthusiastic about them.  I would not say he was the first. He was very enthusiastic and he told me, ``You should look at fiber bundles. They are the right tool for us." I owe this to him. Of course I also  had interactions with Lichnerowicz, but not so much about fiber bundles.
\ids   In the early things that you wrote you do not focus so much on the topological question which I understand was one of the primary motivations for interest in fiber bundles.
\iat  Well,  in my opinion fiber bundles are also a useful tool even when you are considering situations where there is no topology involved. Because otherwise  if you want to describe fields you have to use a preferred coordinate system. This is why I was a little disappointed that Roger was so reluctant.
\ids   Getting back to Roger Penrose just for a moment, he was aware of the Hopf fibration, wasn’t he, of $S^3$?  I thought that you told him about that or he told you. Now I cannot remember precisely what the direction of the communication was.
\iat  It is very hard to say because it is such a long time ago. How did I learn about it? My impression is that I learned about it from mathematicians, not from Roger.
\ids   But then eventually it began to play a very important role in Roger's own thinking?
\iat  Yes. Well – I will think about it. If I find something or if I remember something, then I will let you know but at the present time I do not want to say that it was from me or that it was from Roger.

\ed

\section{International links}

\bd

\ids  I would like to switch direction a little bit now.  Historians of science who are dealing with the history of general relativity are speaking of what they call the Renaissance of General Relativity which is a phenomenon they date to beginning roughly 1955 with the Bern conference and  the subsequent astrophysical discoveries. So that would suggest, however, that if it is a genuine renaissance there must have been some continuity. There must have been some foundational knowledge that was at least being partially recovered and I am wondering if you have heard this notion of renaissance. Is it something that you find to be an accurate description of historical events, dealing with the history of general relativity? Or do you see more of a sense of continuity and growth?
\iat  I would say there is more of a sense of continuity. What happened was that the number of people doing research in relativity has increased very much -  probably because of the  interest in astrophysical problems and discoveries in astrophysics. That is what I think.
\ids  One of the things that these historians maintain is that before this period people were working somewhat in isolation, generally focusing on differing topics and that gradually or even suddenly more international contact arose.
\iat  Yes, but is this specific to relativity? I would say this is really more connected with the general development of science and society.
 Well, I cannot really judge whether this is the case because one would have to look at the number of conferences in various different fields. Well, maybe there is something specific concerning relativity,  because of these astrophysical discoveries.
\ids  Right but certainly the international contact with which you are familiar, I mean first with Infeld and then of course with your own activity -  well in particular with Infeld - that certainly preceded the 1955 meeting which is the date most people identify as the beginning of the Renaissance. So there was some international cooperation already following World War II, I would guess, in the field of relativity. Infeld would be one of the primary examples. But I guess that was not widespread. He did have contact with people with the Synge group in Ireland?
\iat  No, not so much with Ireland, no. 
\ids  But  did he continue relations with Toronto? Or were those relations broken at the time of his departure?
\iat  They were broken. Well the long relation was with Felix Pirani of course.  Felix went to Toronto because of the war.

\ids  Right. Well let’s move forward then to the late 1950s, the early 1960s in which you became very actively involved. I am interested in the role that you personally played in promoting general relativity not just in Poland but also in co-organizing international conferences and helping students to get experience abroad. It seems to me that you played yourself an important role in the development of the international general relativity community. Would you agree?
\iat  Well, no, I do not see myself in this role. Well, of course I would go to conferences and I was – I remember that I was a member of the so to speak non-democratic International Committee on General Relativity and Gravitation.  It was so to speak self-appointed. It had people like Lichnerowicz, Infeld, Fock, Ginzburg, John Wheeler and so on. Well it played a certain role. I remember one thing - that there was a certain conference. There were problems between Israel and the Soviet Union.

\ids Yes, I wanted to ask you specifically about that because I know that certain relativists boycotted the Tbilisi meeting in 1968.

\iat Tbilisi, yes. That's right.

\ids  You remember any details of that controversy that arose at that time?
 I know that Josh Goldberg and Peter Bergmann were involved. They had an exchange of letters with Bondi as I recall. 
 \iat Yes, this is true. Well I remember that I was opposed to boycotting that Tbilisi conference, and I went. That I remember. 
\ids  Would you have been able to boycott it if you had so desired?
\iat  No we wouldn’t have been able to boycott it from Poland. I was against independently of doing this. But  also I was, I remember, against boycotting Israel because there were also such moves to boycott Israel especially from the Soviet side.

In 1971, there was a GRG conference in Copenhagen and a meeting of the International Committee on GRG with a discussion on the next conference planned for Tel-Aviv.
Peter Bergmann, Nathan Rosen, Dmitri Ivanenko, Aleksei Petrov, myself, and several other people, whom I do not remember, were present. There was a vote on the next conference in Israel, a vote by raising hands. The Soviet members of the Committee voted against. I abstained. Bergmann expressed his appreciation for my behavior.

A few months after returning to Poland, I was asked to come to the Rector (President) of Warsaw University, where I was working as professor. There was a mild reprimand: the Rector knew about Copenhagen and my vote.
He said, in such circumstances, I should always vote as the Soviets do.

There is a sequel: a few years after that, I went to another conference (not the important GRG, but something also on gravitation, I think in Eastern Germany). Ivanenko approached me and said something like this: ``I know that you had problems and a reprimand in Warsaw. All this is the fault of Petrov." One should add that Petrov died in 1972.
 
\ed

\bibliographystyle{plain}
\bibliography{qgrav-V19}
\end{document}